\documentclass[reprint,twocolumn]{revtex4-1}
\usepackage{graphicx}
\usepackage{float}
\usepackage{amsmath}
\usepackage{amsfonts}
\usepackage{wasysym}
\usepackage{mathrsfs}
\usepackage{color}
\begin{document}
\title{Oscillatory dynamics of the classical Nonlinear Schrodinger equation}

\author{D.S. Agafontsev$^{(a)}$, V.E. Zakharov$^{(a),(b),(c)}$}
\affiliation{\small \textit{ $^{(a)}$ P. P. Shirshov Institute of Oceanology, 36 Nakhimovsky prosp., Moscow 117218, Russia\\
$^{(b)}$ L. D. Landau Institute for Theoretical Physics, 2 Kosygin str., 119334 Moscow, Russia\\
$^{(c)}$ Department of Mathematics, University of Arizona, Tucson, AZ, 857201, USA}}

\begin{abstract}
We study numerically the statistical properties of the modulation instability (MI) developing from condensate solution seeded by weak, statistically homogeneous in space noise, in the framework of the classical (integrable) one-dimensional Nonlinear Schrodinger (NLS) equation. We demonstrate that in the nonlinear stage of the MI the moments of the solutions amplitudes oscillate with time around their asymptotic values very similar to sinusoidal law. The amplitudes of these oscillations decay with time $t$ as $t^{-3/2}$, the phases contain the nonlinear phase shift that decays as $t^{-1/2}$, and the period of the oscillations is equal to $\pi$. The asymptotic values of the moments correspond to Rayleigh probability density function (PDF) of waves amplitudes appearance. We show that such behavior of the moments is governed by oscillatory-like, decaying with time, fluctuations of the PDF around the Rayleigh PDF; the time dependence of the PDF turns out to be very similar to that of the moments. We study how the 
oscillations that we observe depend on the initial noise properties and demonstrate that they should be visible for a very wide variety of statistical distributions of noise.
\end{abstract}

\maketitle


\section{Introduction.}

The problem of the modulation instability (MI) was first discovered by T.B. Benjamin and J.E. Feir in 1967 for periodic surface gravity waves \cite{benjamin1967disintegration} and since then remains one of the most difficult and interesting problems of mathematical physics. In 1968 V.E. Zakharov \cite{zakharov1968stability} came independently to the same results and demonstrated that the instability observed in \cite{benjamin1967disintegration} using direct surface shape equations was in fact the manifestation of the MI of the condensate solution 
\begin{equation}\label{condensate}
\Psi = Ce^{i\gamma|C|^{2}t},
\end{equation}
for the classical one-dimensional Nonlinear Schrodinger (NLS) equation of focusing type,
\begin{equation}\label{nlse}
i\Psi_t +\beta\Psi_{xx}+\gamma|\Psi|^2 \Psi = 0.
\end{equation}
Here $t$ is time, $x$ is spacial coordinate, $\beta$ and $\gamma$ are real nonzeroth coefficients so that $\beta\gamma>0$, and $\Psi$ is wave field envelope. 

Today the classical NLS equation is recognized as a universal model describing the evolution of the envelope of  quasimonochromatic wave train in weakly nonlinear media \cite{zakharov1984theory}. It has a vast number of applications from surface water waves and propagation of light pulses to Bose-Einstein condensate theory and plasma waves \cite{zakharov1968stability, peregrine1983water, agrawal2007nonlinear, ruprecht1995time, kuznetsov1977solitons}. The evolution of its simplest condensate solution during the MI, however, is still under discussion \cite{zakharov2013nonlinear}. 

Let us suppose that 
$$
\Psi|_{t=0}=C+\epsilon(x)
$$
is the initial condensate state (\ref{condensate}) seeded by small noise $|\epsilon(x)|\ll |C|$. After the scaling and gauge transformations $x=\tilde{x}\sqrt{\beta/(\gamma|C|^{2})}$, $t=\tilde{t}/(\gamma|C|^{2})$, $\Psi=C\tilde{\Psi}e^{i\tilde{t}}$ and $\epsilon=C\tilde{\epsilon}e^{i\tilde{t}}$, the problem of the evolution of this state is reduced to
\begin{equation}\label{Eq01}
i\Psi_t -\Psi +\Psi_{xx}+|\Psi|^2 \Psi = 0,\quad \Psi|_{t=0} = 1+\epsilon(x),
\end{equation}
where all tilde signs are omitted. In terms of Eq. (\ref{Eq01}), the MI develops on the background of the exact condensate solution $\Psi=1$, amplifying small periodic modulations 
$$
\Psi = 1 + \kappa\exp(ikx+i\Omega t),\quad \Omega^{2}=k^{4}-2k^{2},
$$
for wavenumbers $k\in (-\sqrt{2}, \sqrt{2})$, and the maximum increment of the instability $\max_{k}\mathrm{Im}\,\Omega = 1$ is realized at $|k|=k_{0}=1$. When these modulations are small, their evolution can be effectively described by the linearized equations \cite{zakharov1968stability}, and the corresponding stage of the MI is called linear one. As the modulations grow, the linearization no longer works and the full classical NLS equation is necessary. This corresponds to nonlinear stage of the MI. 

In the current publication we demonstrate the oscillatory dynamics of the classical NLS equation in the nonlinear stage of the MI. The first indication for this was in fact obtained in \cite{agafontsev2012rogue}, where the statistical properties of the MI development were studied. Namely, it turned out that the averaged over ensemble of initial data kinetic and potential energy regularly oscillate with time around their asymptotic values, and the amplitude of these oscillations decays with time. The ensemble contained $10^{4}$ realizations of initial data in the form of the condensate state $\Psi=1$ seeded by small noise $\epsilon(x)$; the statistical properties of noise were fixed and the realization of noise varied within the ensemble. The similar behavior was also shown for the probability density function (PDF) of waves amplitudes appearance.

Here we continue this study. We demonstrate that in the nonlinear stage of the MI the moments $M^{(n)}(t)$ of the amplitudes $|\Psi(x,t)|$ with exponents $n\neq 2$ oscillate with time around their asymptotic values $M^{(n)}_{A}$ very similar to sinusoidal law, while the amplitudes of these oscillations decay with time as $t^{-3/2}$. The phases of these oscillations contain the nonlinear phase shift that decays as $t^{-1/2}$, and the period of the oscillations is equal to $\pi$. Under the moments $M^{(n)}(t)$ we understand
\begin{eqnarray}
&&M^{(n)}(t)=\bigg\langle\frac{1}{L}\int_{-L/2}^{+L/2} |\Psi(x,t)|^{n}\,dx\bigg\rangle^{1/n}=\nonumber\\
&&=\bigg(\int_{0}^{+\infty} |\Psi|^{n} P(|\Psi|,t)\,d|\Psi|\bigg)^{1/n},\label{Mn}
\end{eqnarray}
where $n$ is integer, $\langle...\rangle$ stands for arithmetic average over the ensemble of initial data (here and below - ensemble average), $L=\int dx$ is the length of the integration region and $P(|\Psi|,t)$ is the ensemble average PDF to meet amplitude $|\Psi|$ at time $t$. We prove that the asymptotic values $M^{(n)}_{A}$ of the moments $M^{(n)}(t)$, around which the oscillations occur, coincide with the values $M^{(n)}_{R}$ of the moments corresponding to purely Rayleigh PDF $P_{R}(|\Psi|)$.

We demonstrate that such behavior of the moments is governed by the fluctuations of the PDF $P(|\Psi|,t)$ around the Rayleigh PDF $P_{R}(|\Psi|)$; these fluctuations evolve in oscillatory way and decay with time, so that the asymptotic PDF $P(|\Psi|,t)$ coincides with Rayleigh one $P_{R}(|\Psi|)$. The time dependence of the oscillations of the PDF $P(|\Psi|,t)$ turns out to be very similar to that of the moments $M^{(n)}(t)$. Finally, we examine how the oscillations that we observe depend on the statistical properties of the initial noise, and show that they should be visible for a very wide variety of the statistical distributions of noise.

In this publication we do not study the oscillations of the kinetic,
$$
K = \int_{-L/2}^{+L/2}|\Psi_{x}|^{2}\,dx,
$$
and the potential energy, 
$$
U = -\frac{1}{2}\int_{-L/2}^{+L/2}|\Psi|^{4}\,dx,
$$
since the potential energy $U$ corresponds to the moment $M^{(4)}(t)$ as
$$
U = -\frac{L}{2} [M^{(4)}(t)]^{4},
$$
and kinetic energy $K$ oscillates antiphase with the potential one since their sum $E=K+U$ -- the total energy -- is conserved by the classical NLS equation.


\section{Numerical methods.} 

We integrate Eq. (\ref{Eq01}) numerically in the box $-128\pi \le x<128\pi$ with periodic boundary over the period of time $t\in [0, 200]$. We start from the initial data $\Psi|_{t=0}=1+\epsilon(x)$ where $|\epsilon(x)|\ll 1$ is statistically homogeneous in space stochastic noise,
\begin{eqnarray}\label{noise_k_space}
\epsilon(x)=A_{0}\bigg(\frac{L\sqrt{8\pi}}{\theta}\bigg)^{1/2} \int e^{-k^{2}/\theta^{2}+i\xi_{k}+ikx}\,\frac{dk}{2\pi},
\end{eqnarray}
$A_{0}$ is noise amplitude, $L=256\pi$ is the length of the integration region, $\theta$ is noise width in k-space, and $\xi_{k}$ are arbitrary phases for each $k$. The average squared amplitude of noise in x-space can be calculated as,
\begin{eqnarray}
&&\overline{|\epsilon|^{2}}=\frac{\int |\epsilon|^{2}\, dx}{\int dx}=\frac{L\sqrt{8\pi}}{\theta}\frac{A_{0}^{2}}{L}\times \nonumber\\
&&\times\int_{-L/2}^{L/2} e^{-(k_{1}^{2}+k_{2}^{2})/\theta^{2}+i(\xi_{k_{1}}-\xi_{k_{2}})+i(k_{1}-k_{2})x}\, \frac{dk_{1}dk_{2}}{(2\pi)^{2}}dx \approx \nonumber\\ 
&&\approx A_{0}^{2}.\label{noise_x_space}
\end{eqnarray}
Below we will concentrate on the experiment where the ensemble of initial data was generated with noise parameters $A_{0}=10^{-5}$ and $\theta=5$; such noise in the range of the MI $k\in (-\sqrt{2}, \sqrt{2})$ can be treated as a white noise. Then we will demonstrate comparison with the experiments with different ensembles of initial data, corresponding to different values of $A_{0}$ and $\theta$. Note that in \cite{agafontsev2012rogue} the smaller integration region was used, $L=32\pi$. In this publication we had to use $L=256\pi$ because we found that on the period of time $t\in[0, 200]$ the oscillations that we observe depend on $L$ if $L<128\pi$.

We use Runge-Kutta 4th-order method. In order to improve simulations and save computational resources, we employ adaptive change of the spacial grid size $\Delta x$ reducing it when Fourier components of the solution $\Psi_{k}$ at high wave numbers $k$ exceed $10^{-13}\max|\Psi_{k}|$, and increasing $\Delta x$ when this criterion allows. In order to prevent appearance of numerical instabilities, time step $\Delta t$ also changes with $\Delta x$ as $\Delta t = h\Delta x^{2}$, $h \le 0.1$. 

The classical NLS equation (\ref{nlse}) is completely integrable in terms of the inverse scattering method, and has an infinite number of integrals of motion \cite{zakharov1984theory}. The first three of these integrals are wave action, 
\begin{equation}\label{wave_action}
N=\int_{-L/2}^{+L/2} |\Psi(x,t)|^{2}\,dx,
\end{equation}
momentum,
\begin{equation}\label{momentum}
P = \frac{i}{2}\int_{-L/2}^{+L/2}(\Psi_{x}^{*}\Psi - \Psi_{x}\Psi^{*})\,dx,
\end{equation}
and total energy,
\begin{equation}\label{energy}
E = \int_{-L/2}^{+L/2}\bigg(|\Psi_{x}|^{2}-\frac{|\Psi|^{4}}{2}\bigg)\,dx.
\end{equation}
Our scheme of numerical simulations provides very good conservation of the first 12 integrals of motion (see \cite{zakharov1984theory} for more information); for integrals with odd numbers (like wave action and total energy) - with relative accuracy better than $10^{-7}$, and for integrals with even numbers (like momentum) - with absolute accuracy better than $10^{-7}$. We measure absolute errors for integrals of motion with even numbers, since for our initial data these integrals are very close to zero. The first three integrals are conserved by our numerical scheme with accuracy better than $10^{-12}$ (relative error for wave action and total energy, and absolute error for momentum).

For our experiments we use ensembles of 1000 initial distributions each. We checked our statistical results against the size of the ensembles, the parameters of our numerical scheme and the implementation of other numerical methods (Runge-Kutta 5th order and Split-Step 2nd and 4th order methods \cite{muslu2005higher, mclachlan1995numerical}, and also the integration of the classical NLS equation with the help of Ablowitz-Ladik equation with strong coupling between the nodes of the grid \cite{agafontsev2014extreme}), and found no difference.


\section{Oscillatory behavior.} 

\begin{figure}[t] \centering
\includegraphics[width=8cm]{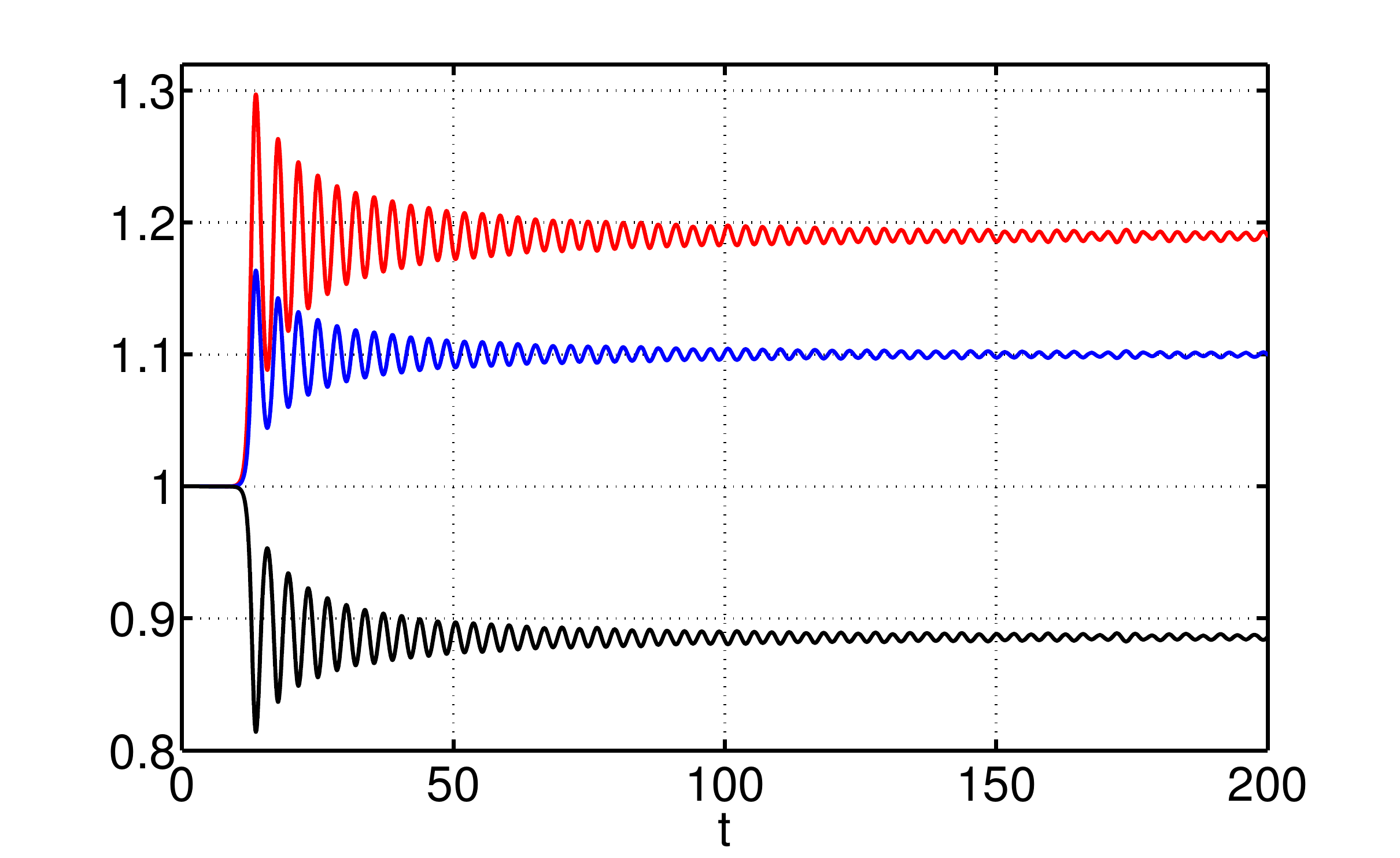}

\caption{\small {\it  (Color on-line) Evolution of the moments $M^{(1)}(t)$ (black), $M^{(3)}(t)$ (blue) and $M^{(4)}(t)$ (red). Ensemble of initial data was generated with noise parameters $A_{0}=10^{-5}$, $\theta=5$.}}
\label{fig:fig1}
\end{figure}

FIG.~\ref{fig:fig1} shows oscillations of the moments $M^{(n)}(t)$ with exponents $n=1$, $n=3$ and $n=4$ for the ensemble of initial data generated with noise parameters $A_{0}=10^{-5}$, $\theta=5$. The moment with exponent $n=2$ does not oscillate because (see (\ref{Mn})) 
\begin{equation}\label{Mn2}
M^{(2)}=\sqrt{\langle N\rangle/L},
\end{equation}
where wave action $N$ (\ref{wave_action}) is conserved by the classical NLS equation, and $\langle N\rangle\approx L$ since $\Psi|_{t=0}=1+\epsilon(x)$, $|\epsilon(x)|\ll 1$; therefore $M^{(2)}(t)\approx 1$. For $n\neq2$ oscillations start in the nonlinear stage of the MI at $t\sim 12$; $M^{(n)}(t)$ for $n\ge3$ oscillate in-phase so that the positions of their minimums and maximums coincide, and antiphase with $M^{(1)}(t)$ so that the positions of minimums of $M^{(1)}(t)$ coincide with the positions of maximums of $M^{(n)}(t)$, $n\ge3$, and vice versa. 

According to Eq. (\ref{Mn}), the oscillations of the moments $M^{(n)}(t)$ must be governed by the fluctuations of the PDF $P(|\Psi|,t)$ to meet amplitude $|\Psi|$ at time $t$. As in \cite{agafontsev2012rogue}, it will be more convenient for us to work with the PDF $P(|\Psi|^{2},t)$ to meet squared amplitude $|\Psi|^{2}$. Since $\int F(x)x\,dx = (1/2)\int F(x)\,d\,x^{2}$, such PDF is connected with the amplitude PDF $P(|\Psi|,t)$ as
$$
P(|\Psi|^{2},t) = \frac{1}{2|\Psi|}P(|\Psi|,t).
$$
Therefore, if the amplitude PDF $P(|\Psi|,t)$ is Rayleigh one,
\begin{equation}\label{Rayleigh}
P_{R}(|\Psi|)=\frac{2|\Psi|}{\sigma^{2}}\exp(-|\Psi|^{2}/\sigma^{2}),
\end{equation}
then the squared amplitude PDF is exponential,
\begin{equation}\label{PDF2}
P_{R}(|\Psi|^{2}) = \frac{1}{\sigma^{2}}\exp(-|\Psi|^{2}/\sigma^{2}),
\end{equation}
and vice versa. We will call squared amplitude PDF (\ref{PDF2}) as Rayleigh one for simplicity.


\begin{figure}[t] \centering
\includegraphics[width=8cm]{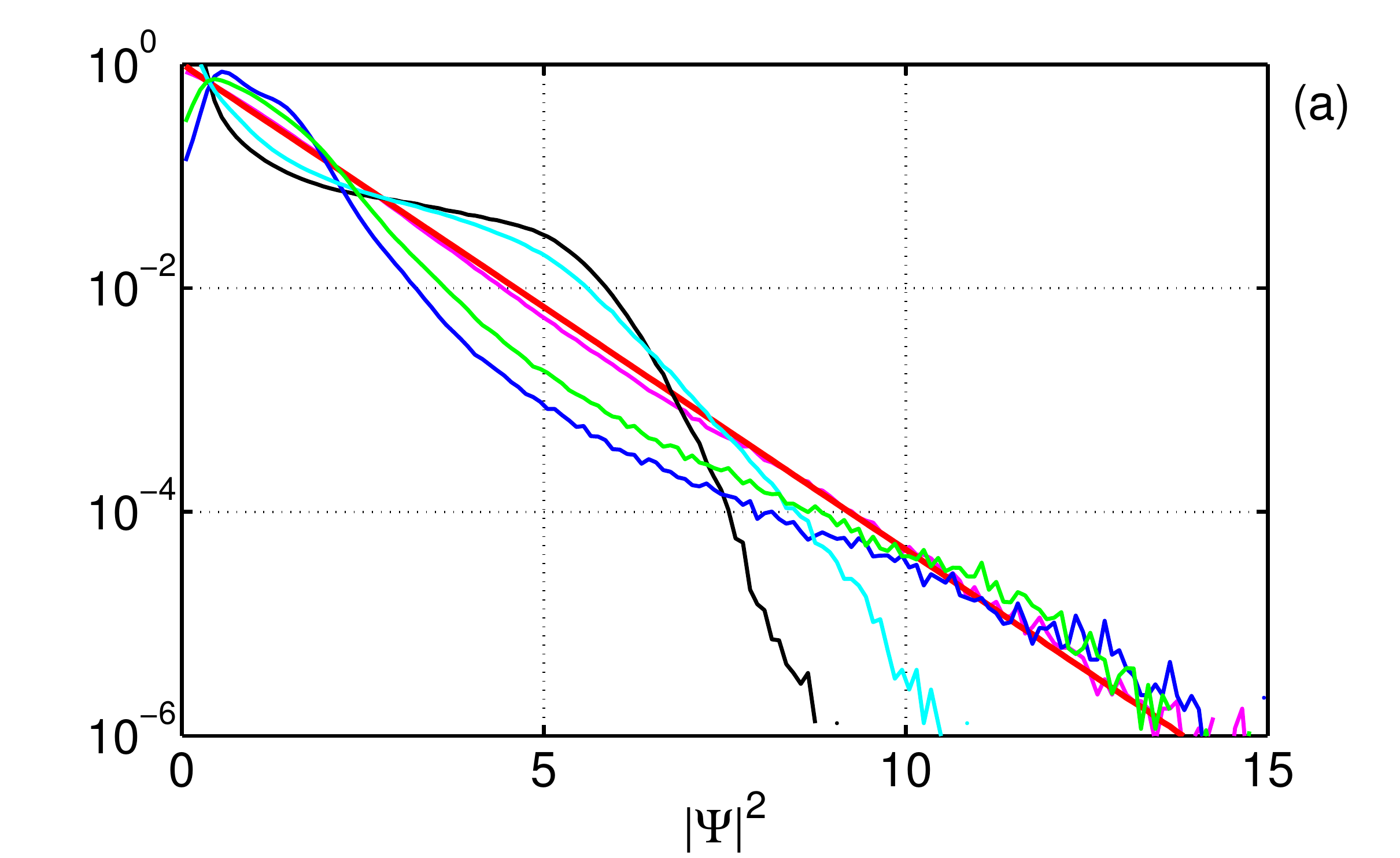}
\includegraphics[width=8cm]{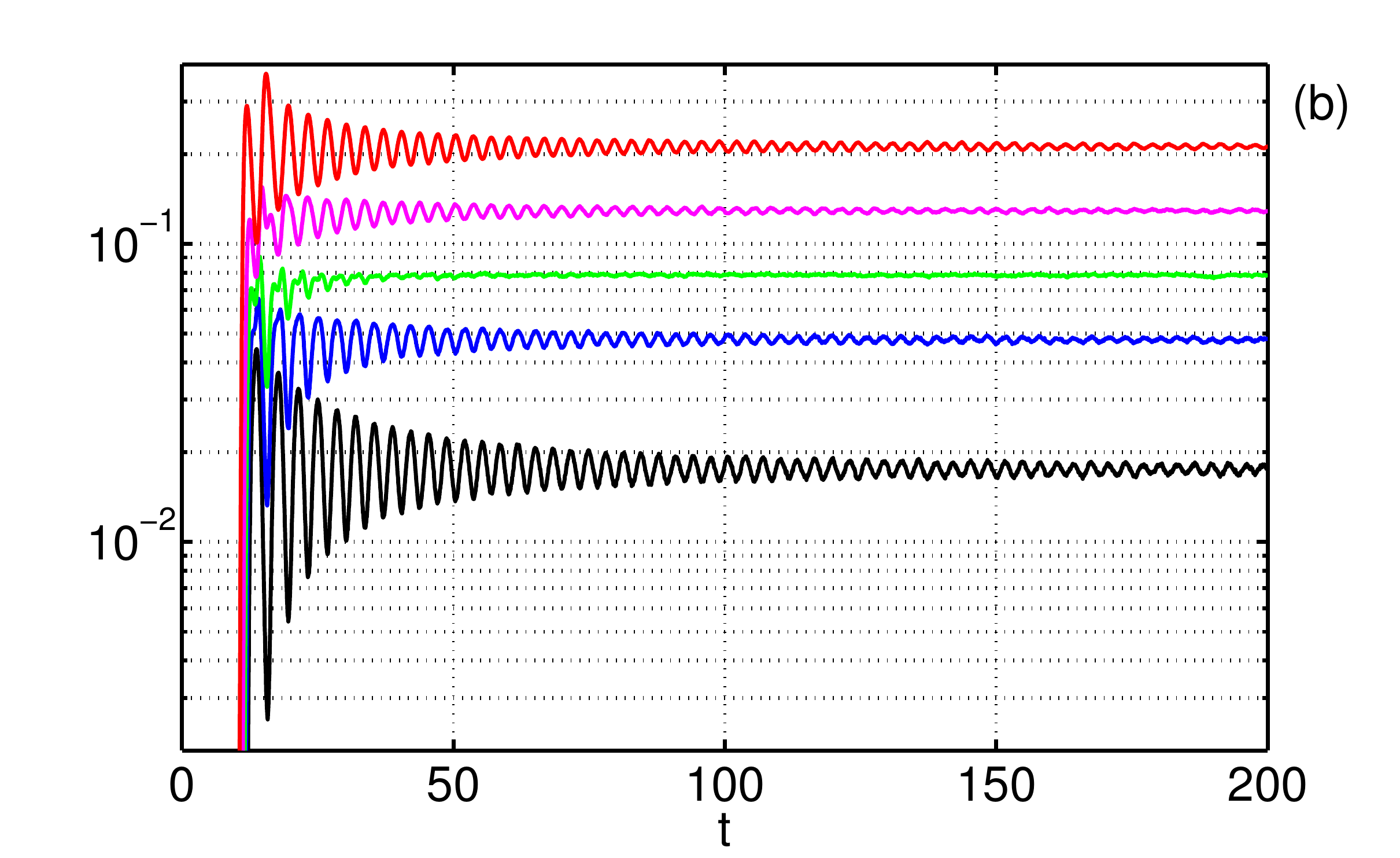}

\caption{\small {\it  (Color on-line) Graph (a): ensemble average PDFs $P(|\Psi|^{2})$ to meet squared amplitude $|\Psi|^{2}$ at the points of time corresponding to local extremums of the moment $M^{(1)}(t)$ -- at $t=13.75$ (black line, local minimum of $M^{(1)}$), $t=15.8$ (blue, maximum), $t=17.8$ (cyan, minimum), $t=19.6$ (green, maximum), and also at $t=100$ (pink), and the asymptotic PDF calculated by averaging the PDF over time $t\in[150, 200]$ (thick red line). The asymptotic PDF coincides with the exponent $\exp(-|\Psi|^{2})$. Graph (b): time dependence of the squared amplitude PDF $P(|\Psi|^{2},t)$ at $|\Psi|^{2}=4$ (black), $|\Psi|^{2}=3$ (blue), $|\Psi|^{2}=2.5$ (green), $|\Psi|^{2}=2$ (pink) and $|\Psi|^{2}=1.5$ (red), log-scale on OY axis. Ensemble of initial data was generated with noise parameters $A_{0}=10^{-5}$, $\theta=5$.}}
\label{fig:fig22}
\end{figure}

\begin{figure}[t] \centering
\includegraphics[width=8cm]{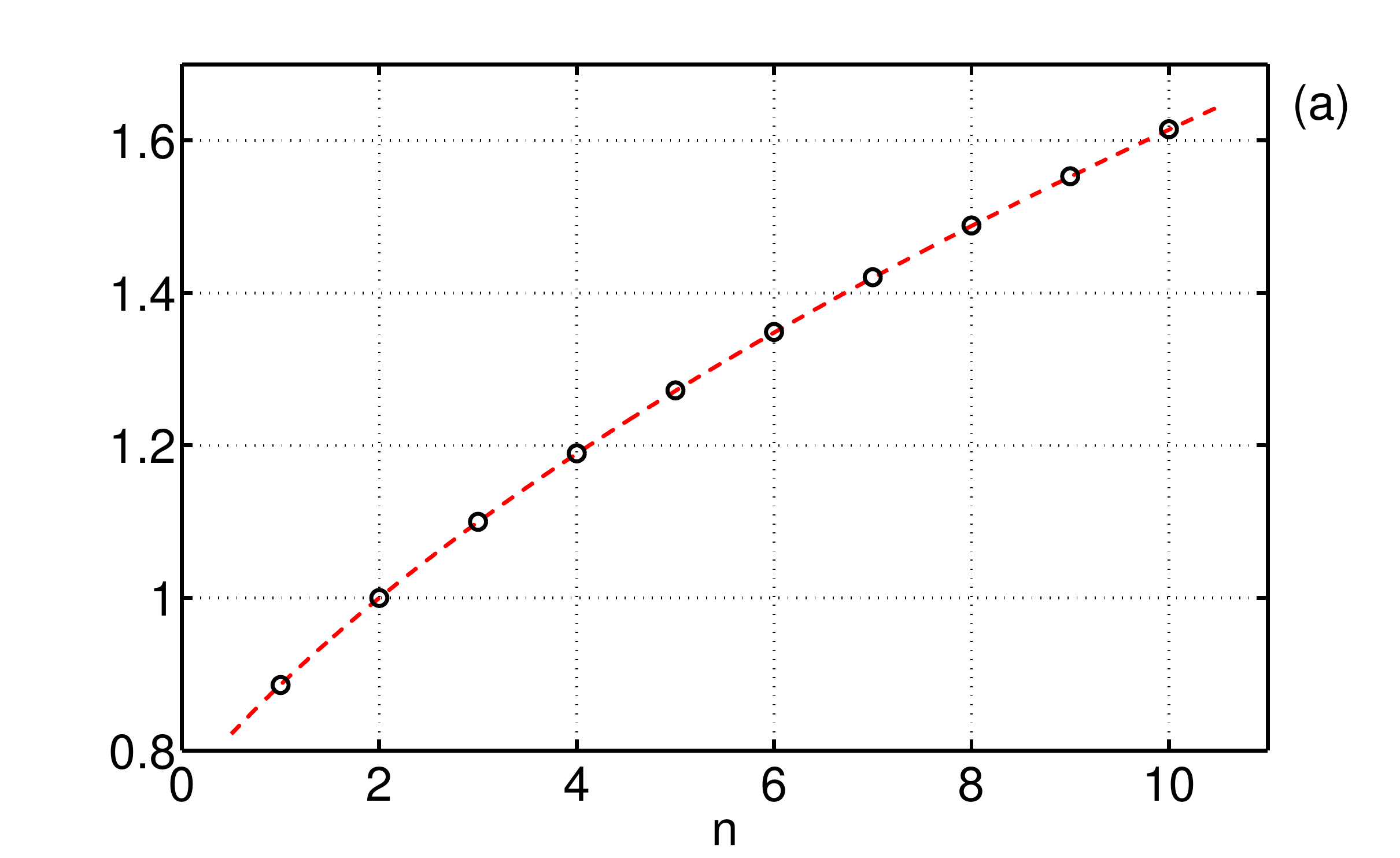}
\includegraphics[width=8cm]{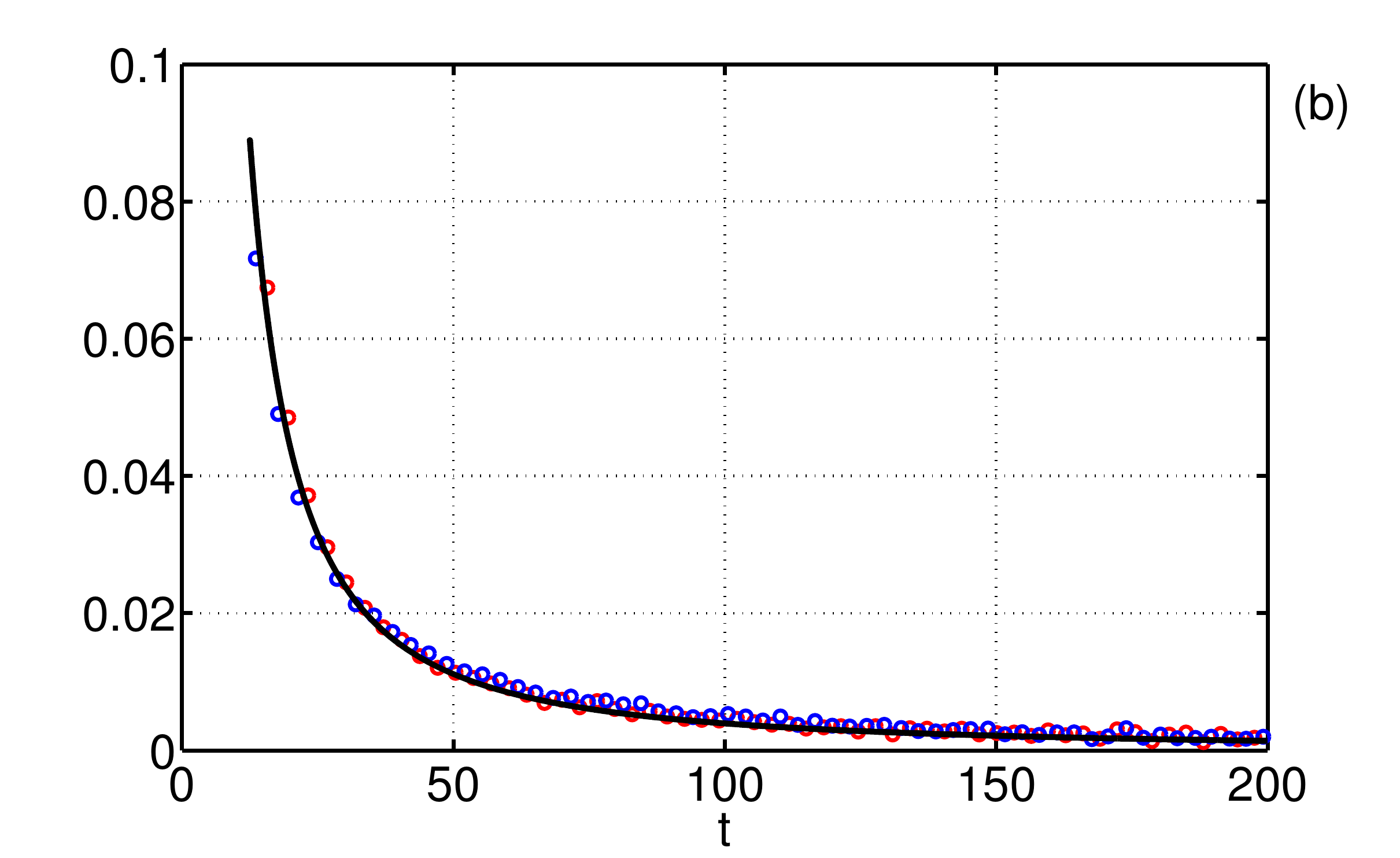}
\includegraphics[width=8cm]{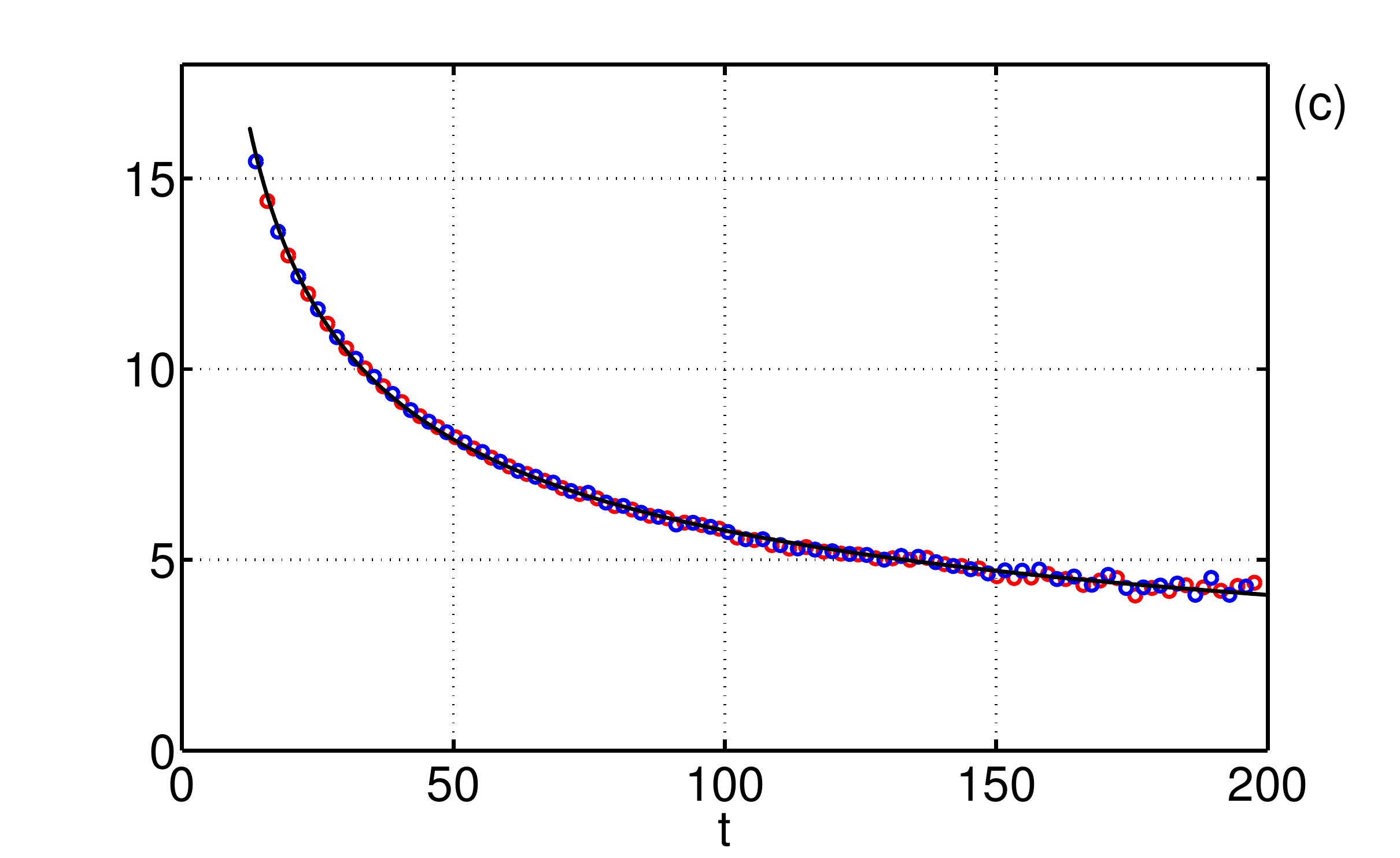}

\caption{\small {\it  (Color on-line) Graph (a): asymptotic values $M^{(n)}_{A}$ (black circles) of the moments $M^{(n)}(t)$, and their Rayleigh approximations $M^{(n)}_{R}$ (\ref{MnR2}) (red dashed line), depending on exponent $n$ for $n=1,...,10$. Graph (b): amplitude of the oscillations of $M^{(1)}(t)$ (circles), calculated as the deviations of the extremums of $M^{(1)}(t)$ from $M^{(1)}_{A}$, depending on time $t$. Graph (c): nonlinear phase shift (\ref{nps}) calculated at the extremums of $M^{(1)}(t)$ (circles), depending on time $t$. On graphs (b) and (c) red circles mark local maximums, and blue circles - local minimums of $M^{(1)}(t)$. Black line on graph (b) is fit by function $a/t^{3/2}$, $a\approx 3.94$, on graph (c) is fit by function $c/\sqrt{t}$, $c\approx 57.7$. Ensemble of initial data was generated with noise parameters $A_{0}=10^{-5}$, $\theta=5$.}}
\label{fig:fig11}
\end{figure}

\begin{figure*}[t] \centering
\includegraphics[width=14cm]{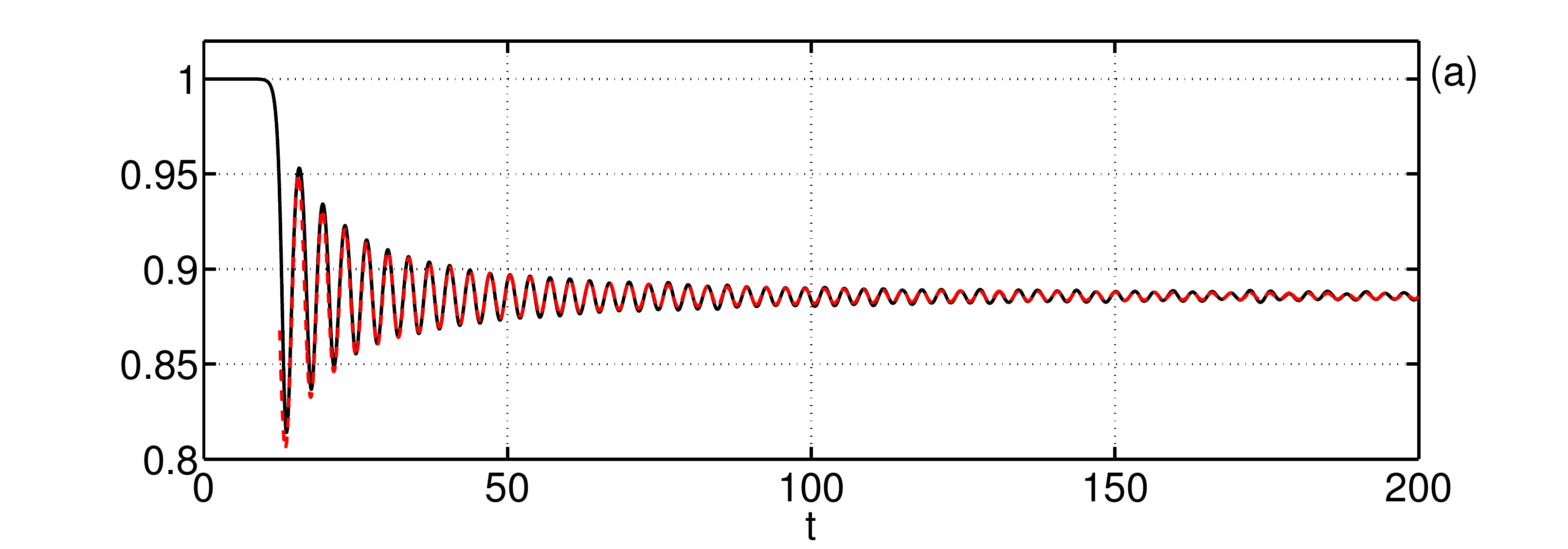}
\includegraphics[width=14cm]{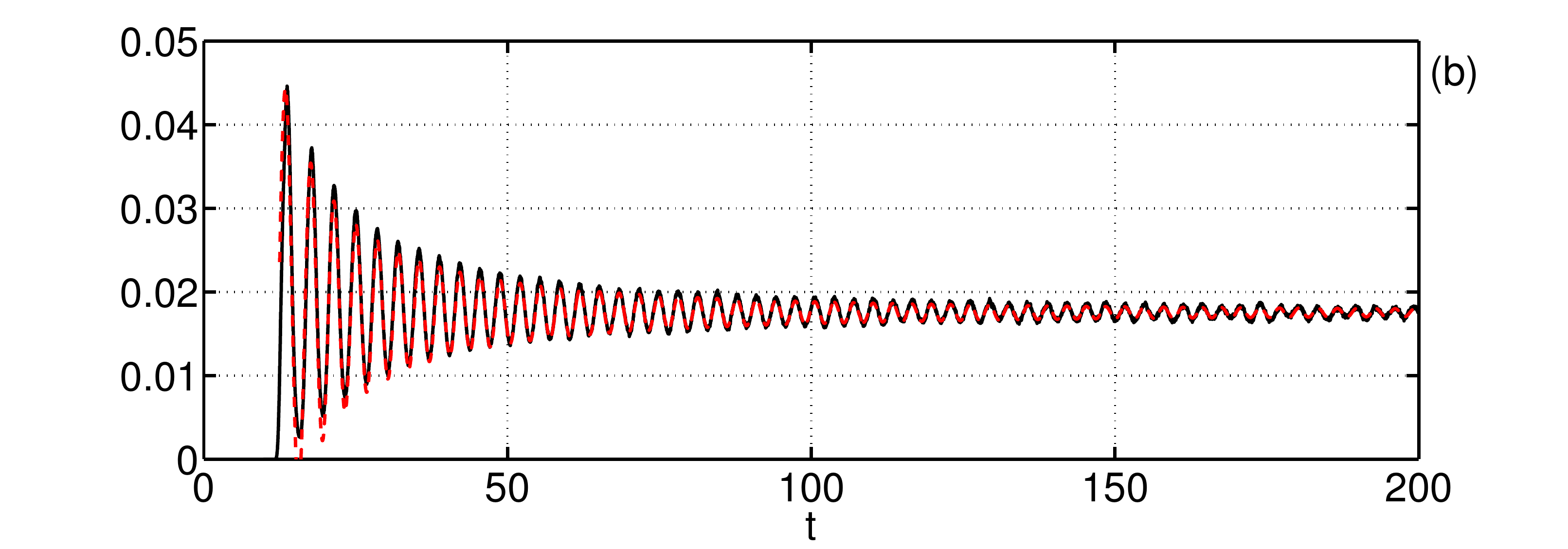}

\caption{\small {\it  (Color on-line) Evolution of the moment $M^{(1)}(t)$ (a) and the squared amplitude PDF $P(|\Psi|^{2},t)$ at $|\Psi|^{2}=4$ (b). Dashed red lines are fits by function $f(t)=f_{0} + [a/t^{3/2}]\sin(bt + c/\sqrt{t} + \phi_{0})$ with parameters $f_{0}=0.886$, $a=3.94$, $b=1.99$, $c=57.7$, $\phi_{0}=-44.1$ for graph (a) and $f_{0}=0.0175$, $a=1.32$, $b=1.99$, $c=57.7$, $\phi_{0}=-44.1+\pi=-41$ for graph (b). Ensemble of initial data was generated with noise parameters $A_{0}=10^{-5}$, $\theta=5$.}}
\label{fig:fig2}
\end{figure*}

\begin{figure*}[t] \centering
\includegraphics[width=14cm]{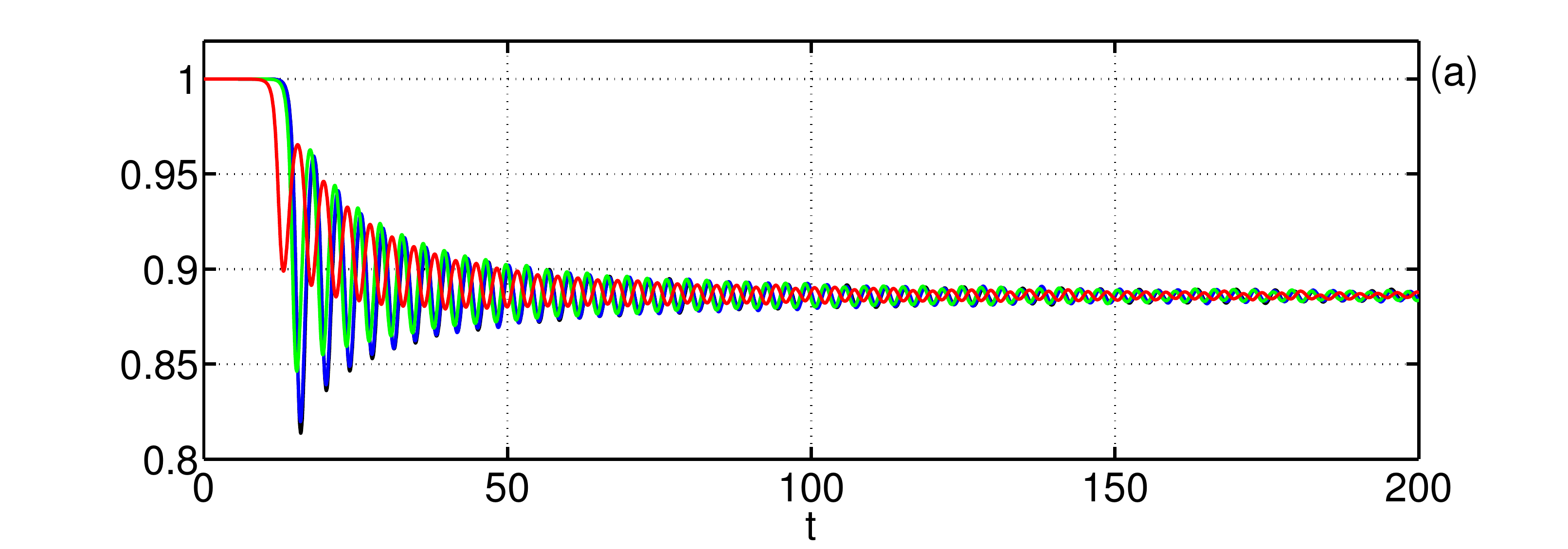}
\includegraphics[width=14cm]{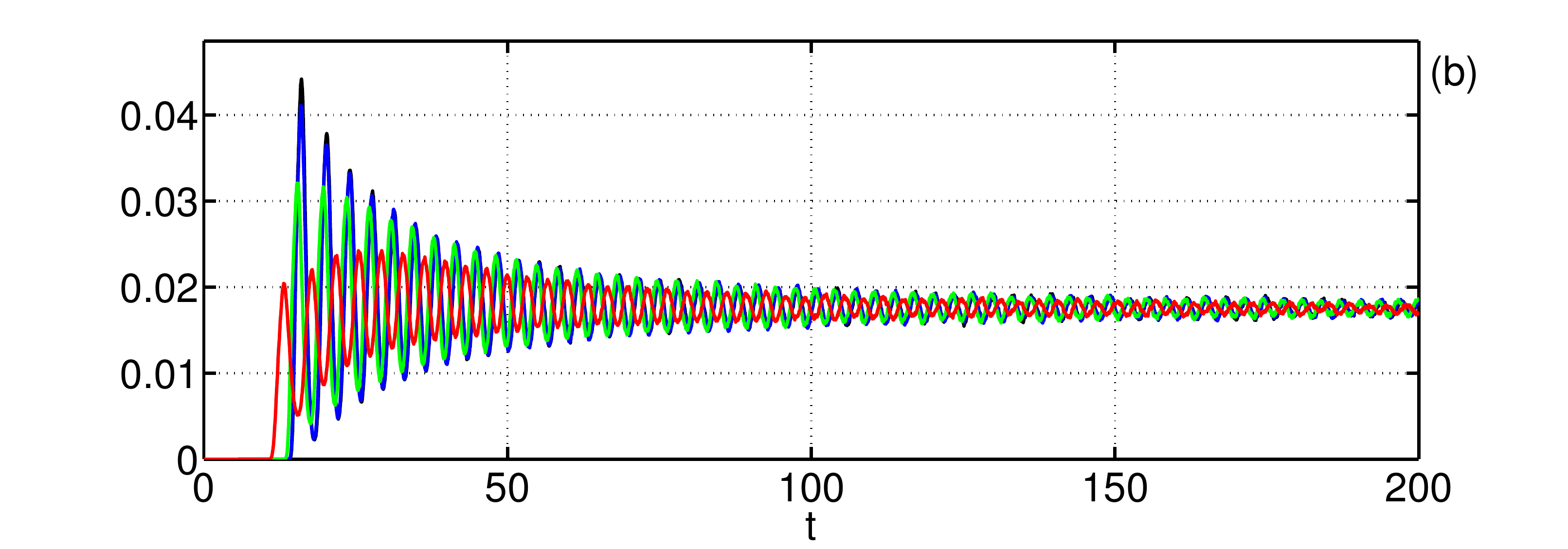}

\caption{\small {\it (Color on-line) Evolution of the moment $M^{(1)}(t)$ (a) and the squared amplitude PDF $P(|\Psi|^{2},t)$ at $|\Psi|^{2}=4$ (b) for noise parameters $A_{0}=10^{-6}$, $\theta=5$ (black), $A_{0}\approx 1.17\times 10^{-6}$, $\theta=1$ (blue), $A_{0}\approx 1.66\times 10^{-5}$, $\theta=0.5$ (green) and $A_{0}\approx 8.92\times 10^{-4}$, $\theta=0.35$ (red). During these experiments noise amplitude in k-space at $|k|=1$ (\ref{noise_k_amplitude}) was fixed to $q_{0}\approx 2.73\times 10^{-5}$.}}
\label{fig:fig3}
\end{figure*}


We measure the squared amplitude PDF $P(|\Psi|^{2},t)$ as 
\begin{equation}\label{PDF}
P(|\Psi|^{2},t) = \frac{W(|\Psi|^{2},t)-W(|\Psi|^{2}+\Delta A,t)}{\Delta A},
\end{equation}
where $W(Y,t)$ is the ensemble average probability to meet squared amplitude $|\Psi|^{2}$ larger than $Y$, and $\Delta A=0.1$ is the bin size. As in \cite{agafontsev2012rogue}, we find that the asymptotic PDF, that we calculate by averaging the PDF over time $t\in [150, 200]$, coincides with the Rayleigh one (\ref{PDF2}) with $\sigma=1$. The evolution of the PDF $P(|\Psi|^{2},t)$ resembles oscillations around the asymptotic (Rayleigh) PDF; the maximum deviations from Rayleigh PDF are achieved at the points of time corresponding to local maximums and local minimums of the moments $M^{(n)}(t)$. The examples of such deviations are shown on FIG.~\ref{fig:fig22}a for two local maximums and two local minimums of $M^{(1)}(t)$; the deviations corresponding to maximums of $M^{(1)}(t)$ are close to opposite with respect to the deviations corresponding to minimums of $M^{(1)}(t)$.

The fluctuations of the PDF gradually decay with time, so that starting from $t\sim 100$ the PDF is nearly indistinguishable from Rayleigh one (\ref{PDF2}) with $\sigma=1$ (see FIG.~\ref{fig:fig22}a). FIG.~\ref{fig:fig22}b shows, in logarithmic scale on OY axis, the time dependence of the PDF $P(|\Psi|^{2},t)$ at fixed points of squared amplitude $|\Psi|^{2}$. For $|\Psi|^{2}\in (0.5, 2)$ the PDF oscillates in-phase, and for $|\Psi|^{2}\in (3, 6)$ -- antiphase with $M^{(1)}(t)$; the amplitudes of these oscillations decay with time. It is difficult to determine the time dependence of the PDF beyond these two regions of the squared amplitude, since in addition to oscillations the PDF changes its shape from very high and thin peak around $|\Psi|^{2}=1$ in the linear stage of the MI to Rayleigh PDF (\ref{PDF2}) with $\sigma=1$ at late times $t\sim 100$.

Below we study the time dependence of the oscillations in details on the example of the moment $M^{(1)}(t)$. Moments $M^{(n)}(t)$ with exponents $n\ge 3$, as well as the PDF $P(|\Psi|^{2},t)$ at $|\Psi|^{2}\in (0.5, 2)$ and $|\Psi|^{2}\in (3, 6)$, oscillate according to the same law as $M^{(1)}(t)$. The PDF $P(|\Psi|^{2},t)$ at $|\Psi|^{2}\in (0.5, 2)$ oscillates with the same phase as $M^{(1)}(t)$, while $M^{(n)}(t)$ for $n\ge 3$ and $P(|\Psi|^{2},t)$ at $|\Psi|^{2}\in (3, 6)$ -- with the opposite phase as $M^{(1)}(t)$.

Since the asymptotic PDF coincides with Rayleigh PDF (\ref{PDF2}), the asymptotic values $M_{A}^{(n)}$ of the moments $M^{(n)}(t)$ must coincide with the values of the moments $M_{R}^{(n)}$ that correspond to Rayleigh PDF. The latter can be easily calculated. Indeed, substitution (\ref{Rayleigh}) into (\ref{Mn}) yields
\begin{equation}\label{MnR1}
M_{R}^{(n)}=\bigg(\frac{2}{\sigma^{2}}\int_{0}^{+\infty}|\Psi|^{n+1}e^{-|\Psi|^{2}/\sigma^{2}}\,d|\Psi|\bigg)^{1/n},
\end{equation}
that for $n=2$ gives $M_{R}^{(2)}=\sigma$. From the other hand, since wave action (\ref{wave_action}) is conserved by the classical NLS equation, substitution of the initial data $\Psi|_{t=0}=1+\epsilon(x)$, $|\epsilon(x)|\ll 1$, into (\ref{Mn2}) gives $M_{R}^{(2)}\approx 1$. Therefore, the value of the parameter $\sigma\approx 1$, that we observe for the asymptotic Rayleigh PDF (\ref{PDF2}) in our experimental data, actually follows from the conditions of the conservation of wave action and the form of the initial data.

The condition $\sigma=1$ leads to
\begin{equation}\label{MnR2}
M_{R}^{(n)}=\bigg[\Gamma\bigg(\frac{n}{2}+1\bigg)\bigg]^{1/n},
\end{equation}
where $\Gamma(m)$ is gamma-function,
$$
\Gamma(m)=\int_{0}^{+\infty}x^{m-1}e^{-x}dx.
$$
FIG.~\ref{fig:fig11}a shows that the asymptotic values of the moments $M^{(n)}_{A}$, that we calculated by averaging the moments $M^{(n)}(t)$ over time $t\in[150, 200]$, indeed coincide with the values $M^{(n)}_{R}$ corresponding to Rayleigh PDF (\ref{Rayleigh}) with $\sigma=1$. The asymptotic value of the moment $M^{(4)}_{A}\approx 2^{1/4}$ allows one to calculate the asymptotic values of the ensemble average kinetic $\langle K\rangle/L$ and potential $\langle U\rangle/L$ energy densities as
\begin{equation}\label{kinetic_potential}
\langle K\rangle/L \approx 0.5,\quad\quad \langle U\rangle/L \approx -1.
\end{equation}
Here we use the conservation of the total energy $E=K+U$ and the form of the initial data $\Psi|_{t=0}=1+\epsilon(x)$, $|\epsilon(x)|\ll 1$, that yields $\langle K+U\rangle/L\approx -0.5$. The relation (\ref{kinetic_potential}) means that the turbulence in our system is not weak.

The amplitudes of the oscillations of the moments, that we measure as the deviations of local maximums and local minimums of $M^{(n)}(t)$ from the corresponding asymptotic values $M^{(n)}_{A}$, decay with time as $a/t^{3/2}$, as shown on FIG.~\ref{fig:fig11}b. For our experiment with noise parameters $A_{0}=10^{-5}$, $\theta=5$, the prefactor is equal to $a=(3.94\pm 0.03)$ for $M^{(1)}(t)$. 

It turns out that the period of our oscillations changes from $\Delta T\sim 4$ at $t\sim 20$ to $\Delta T\sim 3$ at $t\sim 200$. We think that this is the effect analogous to the nonlinear phase shift. Indeed, one can search for the approximation of $M^{(n)}(t)$ in the form
\begin{equation}\label{f1}
f(t)=M^{(n)}_{A} + \frac{a}{t^{3/2}}\sin(bt+\phi_{nl}(t)+\phi_{0}),
\end{equation}
where the nonlinear phase shift $\phi_{nl}(t)$ should be proportional to the amplitude of the oscillations $a/t^{3/2}$ multiplied by time $t$, or $\phi_{nl}(t)=c/\sqrt{t}$ where $c$ is constant. Then, the phases $\Phi$ for the local maximums $t_{max}$ of $M^{(n)}(t)$ should be equal to 
$$
\Phi(t_{max}) = bt_{max}+\frac{c}{\sqrt{t_{max}}}+\phi_{0} = \frac{\pi}{2}+2\pi m,
$$
and for the local minimums $t_{min}$ -- to
$$
\Phi(t_{min}) = bt_{min}+\frac{c}{\sqrt{t_{min}}}+\phi_{0} = \frac{3\pi}{2}+2\pi m,
$$
where $m$ is integer number. We find all the subsequent extremums $t_{max}$ and $t_{min}$ of $M^{(n)}(t)$ from one hand, and their phases $\Phi$ from the other hand by setting $m=0$ for the first maximum, $m=1$ for the second maximum, and so on. Then, with the help of the least squares method we determine the coefficients $b$, $c$ and $\phi_{0}$, that in case of $M^{(1)}(t)$ are equal to $b=1.99$, $c=57.7$ and $\phi_{0}=-44.1$. After that we check that the nonlinear phase shift
\begin{equation}\label{nps}
\Phi(t)-bt-\phi_{0},
\end{equation}
calculated at the extremums of $M^{(1)}(t)$, indeed is very well approximated by the function $c/\sqrt{t}$, as shown on FIG.~\ref{fig:fig11}c. 

In our experiments we observe that the anzats (\ref{f1}) fits remarkably well to the experimental data for all moments $M^{(n)}(t)$, $n\neq 2$, that we measure, and also for the PDF $P(|\Psi|^{2},t)$ at $|\Psi|^{2}\in(0.5, 2)$ and $|\Psi|^{2}\in(3, 6)$. The example of such fit for $M^{(1)}(t)$ and for the PDF $P(|\Psi|^{2},t)$ at $|\Psi|^{2}=4$ is shown on FIG.~\ref{fig:fig2}. The absence of the nonlinear phase shift, or the nonlinear phase shift with the exponent significantly different from --0.5, leads to the situation when the anzats (\ref{f1}) does not fit to the oscillations, or fits significantly worse. It is interesting to note that the period of our oscillations $2\pi/b\approx 3.16$ is almost equal to $\pi$. We think that it should coincide with $\pi$, and we measure the same period for all of our experiments.


\section{Dependence on statistics of initial noise.} 

We repeated our experiments for ensembles of initial data with different noise parameters. We didn't find significant dependence of our results on noise amplitude $A_{0}$, except that with decreasing of $A_{0}$ the time necessary for the nonlinear stage of the MI to arrive increases. The period of the oscillations $\Delta T=2\pi/b\approx\pi$ does not depend on $A_{0}$, and in the beginning of the nonlinear stage of the MI the amplitudes of the oscillations are roughly the same for all of our experiments from $A_{0}=10^{-12}$ to $A_{0}=10^{-3}$.

We also tested the following noise distribution, 
\begin{eqnarray}
&&\epsilon_{2}(x)=A_{0}\bigg(\frac{L\sqrt{8\pi}}{\theta}\bigg)^{1/2}\times\nonumber\\ 
&&\times\int 10^{-p_{k}} e^{-k^{2}/\theta^{2}+i\xi_{k}+ikx}\,\frac{dk}{2\pi},\label{noise_k_space2}
\end{eqnarray}
where $p_{k}$ is uniformly distributed over $[0, 10]$ random value for each $k$, $A_{0}=10^{-5}$ and $\theta=5$. The multiplier $10^{-p_{k}}$ introduces the detuning between the amplitudes of noise in $k$-space by up to 10 orders of magnitude. However, we came to very similar results, though the oscillations became slightly less regular. In our opinion this means that the oscillations that we observe should be visible for a very wide variety of statistical distributions of noise.

The most significant dependence of our results was found on the noise width in k-space $\theta$. We performed four experiments changing $\theta$ from $\theta=5$ to $\theta=0.35$. During these experiments we fixed noise amplitude in k-space at $|k|=1$ to 
\begin{equation}\label{noise_k_amplitude}
q_{0} = A_{0}\bigg(\frac{L\sqrt{8\pi}}{\theta}\bigg)^{1/2}\exp(-1/\theta^{2})\approx 2.73\times 10^{-5},
\end{equation}
so that $A_{0}=10^{-6}$ for $\theta=5$. As $|k|=1$ is the fastest growing mode in the linear stage of the MI, and assuming that at the start of the nonlinear stage this mode is the leading one after the zeroth harmonic $k=0$, such condition should provide us the same starting time for the nonlinear stage of the MI for the experiments with different $\theta$. This supposition turns out to be valid for the experiment with $A_{0}=1.17\times 10^{-6}$, $\theta=1$, when all of the functions that we measure, including the moments, the PDFs, the energy spectrum and the spacial correlation functions (see \cite{agafontsev2012rogue} for more information), almost coincide with that for the experiment with noise parameters $A_{0}=10^{-6}$, $\theta=5$. However, for the experiments with $A_{0}=1.66\times 10^{-5}$, $\theta=0.5$ and $A_{0}=8.92\times 10^{-4}$, $\theta=0.35$ we observe significantly different results.

As shown on FIG.~\ref{fig:fig3}, the oscillations become non-symmetric with respect to the asymptotic values of the oscillating functions; however, these asymptotic values still coincide with the Rayleigh approximations. The amplitudes of the oscillations become smaller, and also the local maximums and the local minimums of the oscillating functions belong now to different time dependencies. However, for local maximums of $M^{(1)}(t)$ and $P(|\Psi|^{2},t)$ at $|\Psi|^{2}\in(0.5, 2)$, and local minimums of $M^{(n)}(t)$, $n\ge 3$, and $P(|\Psi|^{2},t)$ at $|\Psi|^{2}\in(3,6)$, we still observe the decaying amplitudes of the oscillations as $\sim\,t^{-3/2}$ and the decaying nonlinear phase shift as $\sim\,t^{-1/2}$. The period of the oscillations $\Delta T=2\pi/b\approx\pi$ does not change with $\theta$.


\section{Acknowledgements.}  

D. Agafontsev thanks E. Kuznetsov for valuable discussions concerning the subject of this publication, M. Fedoruk for access to and V. Kalyuzhny for assistance with Novosibirsk Supercomputer Center. This work was done in the framework of Russian Federation Government Grant (contract No. 11.G34.31.0035 with Ministry of Education and Science of RF), and also supported by the program of Presidium of RAS "Fundamental problems of nonlinear dynamics in mathematical and physical sciences", program of support for leading scientific schools of Russian Federation, RFBR grants 12-01-00943-a, 13-01-00261 and also Sergei Badulin RFBR grant 11-05-01114-a.


\begin{thebibliography}{99}

\bibitem{benjamin1967disintegration} T.B. Benjamin, J.E. Feir, \textit{The disintegration of wave trains on deep water Part 1. Theory}, J. Fluid Mech., vol. 27, part 3, pp. 417-430 (1967).

\bibitem{zakharov1968stability} V.E. Zakharov, \textit{Stability of periodic waves of finite amplitude on the surface of a deep fluid}, Zh. Prikl. Mekh. Tekh. Fiz. 9, 86-94 (1968) [J. Appl. Mech. Tech. Phys. 9, 190-194 (1968)]. 

\bibitem{zakharov1984theory} V.E. Zakharov, S.V. Manakov, S.P. Novikov, L.P. Pitaevsky, \textit{Theory of solitons. The method of the inverse scattering problem}, Consultants Bureau, New York (1984). 

\bibitem{peregrine1983water} D.H. Peregrine, \textit{Water waves, nonlinear Schrodinger equations and their solutions}, J. of the Australian Math. Soc. B, vol. 25, iss. 01, pp. 16-43 (1983).

\bibitem{agrawal2007nonlinear} G.P. Agrawal, P.L. Kelley, I.P. Kaminow, \textit{Nonlinear Fiber Optics}, 3rd ed. Academic, San Diego (2001).

\bibitem{ruprecht1995time} P.A. Ruprecht, M.J. Holland, K. Burnett, M. Edwards, \textit{Time-dependent solution of the nonlinear Schrodinger equation for Bose-condensed trapped neutral atoms}, Phys. Rev. A 51, 4704 (1995).

\bibitem{kuznetsov1977solitons} E.A. Kuznetsov, \textit{Solitons in a parametrically unstable plasma}, Sov.Phys. - Dokl. (Engl. Transl.) vol. 22, 507 (1977) [\textit{On solitons in parametrically unstable plasma}, Doklady USSR (in Russian) vol. 236, 575 (1977)].

\bibitem{zakharov2013nonlinear} V.E. Zakharov, A.A. Gelash, \textit{Nonlinear Stage of Modulation Instability}, Phys. Rev. Lett. 111, 054101 (2013).

\bibitem{agafontsev2012rogue} D.S. Agafontsev, V.E. Zakharov, \textit{Rogue waves statistics in the framework of one-dimensional Generalized Nonlinear Schrodinger Equation}, arXiv:1202.5763v3 (2012).

\bibitem{muslu2005higher} G.M. Muslu, H.A. Erbay, \textit{Higher-order split-step Fourier schemes for the generalized nonlinear Schrodinger equation}, Mathematics and Computers in Simulation, v. 67, iss. 6 (2005).

\bibitem{mclachlan1995numerical} R.I. Mclachlan, \textit{On the numerical integration of ordinary differential equations by symmetric composition methods}, SIAM Journal on Scientific Computing, v. 16, iss. 1 (1995).

\bibitem{agafontsev2014extreme} D.S. Agafontsev, \textit{Extreme waves statistics for Ablowitz-Ladik system}, arXiv:1310.4406 (2013), JETP Letters, vol. 98, issue 11, pp. 826-829 (2013).

\end{thebibliography}
\end{document}